\documentclass[twocolumn,aps]{revtex4}%
\usepackage{graphicx}
\usepackage{amsmath}
\usepackage{amsfonts}
\usepackage{amssymb}%
\begin{document}

\title{Bright cavity polariton solitons}
\author{O.A. Egorov$^{1}$, D.V. Skryabin$^{2}$, A.V. Yulin$^{2}$ and F. Lederer$^{1}$}
\date{\today}
\address{
$^1$Institute of Condensed Matter Theory and Solid State Optics,
Friedrich-Schiller-Universit\"at Jena, Max-Wien-Platz 1, 07743 Jena,
Germany\\
$^2$Centre for Photonics and Photonic Materials, Department of Physics, University of
Bath, Bath BA2 7AY, United Kingdom}

\begin{abstract}
The lower branch of the dispersion relation of exciton polaritons in
semiconductor microcavities, operating in the strong-coupling
regime, contains sections of both positive and negative curvature
along one spatial direction. We show that this  leads to the
existence of stable one-dimensional bright microcavity solitons
supported by the repulsive polariton nonlinearity.  To achieve
localization along the second transverse direction we propose to
create a special soliton waveguide by changing the cavity detuning
and hence the boundary of the soliton existence in such a way that
the solitons are allowed only within the stripe of the desired
width.
\end{abstract}


\maketitle

\input{epsf.tex} \epsfverbosetrue

\narrowtext
The study of nonlinear and quantum effects in optical
microcavities is an area of current vigorous activity
\cite{Deveaud2007,Kavokin2008,Peschel2003}. In particular, the
strong coupling regime in quantum well semiconductor microcavities
\cite{Deveaud2007,Kavokin2008} leads to a strong and fast nonlinear
response of microcavity exciton polaritons, which have been used for
low threshold bistability
\cite{Tredicucci1996,Baas2004,Carusotto2004,cnrs}, parametric wave
mixing
\cite{Houdre2000,Savvidis2000,Gippius2004,Diederichs2006,Wouters07,Krizhanovskii2008}
and for the demonstration of long range polariton coherence
supporting claims for observation of Bose-Einstein condensation
\cite{Stevenson2000,Deng2002,Kasprzak2006,Balili2007,Love2008,Baas2008}.
Research in this area bordering condensed matter physics and
photonics has  links with many  topics of current interest, which
include, but not limited to, superfluidity, solitons, vortices,
photonic crystals and quantum information.

Diffraction and dispersion are important features of waves of any
nature  and the problem of their  control  stimulates significant
research efforts. The recent time has witnessed remarkable
developments into photonic crystals, which can be used to fully
compensate diffraction and to control its value and sign, see, e.g.,
\cite{Eisenberg2000,Pertsch2002}. The ideas developed for
electromagnetic waves have also been applied to matter waves
\cite{Greiner2002}, surface plasmons \cite{Ebbesen1998} and other
waves. Microcavity polaritons have  been a part of these  research
efforts, with periodic potentials for polaritons created by the
mirror patterning or by the surface acoustic waves
\cite{Lai2007,Cho2005}. Note, that the term diffraction is  most
commonly used when the beam spreading happens upon propagation in
space. While, in the context of planar microcavities one deals with
the temporal evolution of spots of light, which is analogous to the
spreading of quantum wave packets. This evolution is governed by the
dispersion law linking the frequency (or energy)
 to the transverse wavenumber (or momentum).

\begin{figure}[ptb]
\setlength{\epsfxsize}{2.5in}
\centerline{\epsfbox{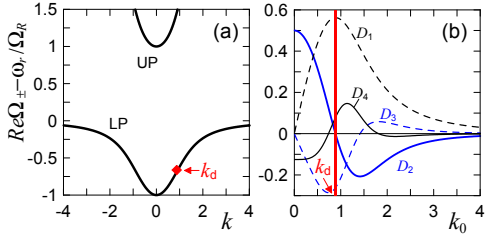}}\caption{(color online) (a) Polariton
dispersion:  lower (LP) and upper(UP) branch in the strong coupling
regime. (b) Dispersion
coefficients vs. inclination of the holding beam $k_{0}$.}%
\label{disp_ch}%
\end{figure}

Diffraction and dispersion control has been comprehensively
discussed and used in conjunction with spatial and temporal soliton
formation, see e.g. \cite{Fleischer2003,Kivshar2001,Lederer2008}. A
particular field of these studies concerns spatial cavity solitons
in the weak coupling regime with periodic modulation of the detuning
and/or other cavity parameters, see, e.g., \cite{Staliunas2003}.
Weak and strong coupling regime differ mainly in that in the latter
case the dispersion curve exhibits two branches signalling mixing of
light and matter excitations. Namely, additionally to the upper
parabolic branch, which is largely a photonic one,  the lower
polariton branch appears. In our context the most relevant feature
is that this lower branch exhibits an inflection point where the
second order dispersion  changes sign [Fig.\ref{disp_ch}(a)]. It is
worth noting that this effect is merely evoked by the strong
photon-exciton coupling, hence it appears even in a homogeneous
cavity and does not require any modulation as in the weak coupling
regime. The primary goal of this Letter is to demonstrate how this
peculiarity can be exploited for the formation of stable
\emph{bright} cavity polariton solitons (CPSs) supported by the
repulsive exciton-exciton interaction. However, the inflection
 of the dispersion  curve provides only localization in
one dimension and additional measures for soliton trapping in the
other dimension have to be implemented relying on the soliton
existence domain shift by means of the space dependent detuning
control.

 Recently it has been suggested to use bright
cavity solitons in the \emph{weak} coupling regime in all optical
processing schemes \cite{Brambilla1997,Barland2002,Pedaci2008}. In
this context bright CPSs in the \emph{strong} coupling regime are
superior and promising candidates for a practical implementation
because they have a picosecond excitation time and require extremely
low pump intensities of about $100$ W/cm$^{2}$, thus surpassing
their weakly coupled counterparts by  almost  two orders of
magnitude on these two crucial parameters \cite{Yulin}.
 The first experimental observations of dark and
bright soliton-like structures in the strong coupling regime of
semiconductor microcavities have been recently reported in
\cite{Larionova2008} calling for more theoretical and experimental
studies of this topic. It has been demonstrated  \cite{Yulin} that
the dark solitons are the only stable solutions with momenta
centered around  the bottom of the lower polariton branch.

The widely accepted dimensionless model for excitons strongly coupled to
cavity photons is \cite{Deveaud2007,Kavokin2008,Carusotto2004,Yulin}
\begin{align}
&  \partial_{t}E-i(\partial_{x}^{2}+\partial_{y}^{2})E+\left[  \gamma
_{c}-i\Delta-iU(y)\right]  E=\nonumber\\
&  =i\Psi+E_0e^{ik_{0}x},\\
&  \partial_{t}\Psi+(\gamma_{0}-i\Delta)\Psi+i|\Psi|^{2}\Psi=iE.
\nonumber\label{e1}%
\end{align}
Here $E$ and $\Psi$ are the averages of the photon and exciton creation or
annihilation operators and polarization effects are disregarded. The
normalization is such that $(\Omega_{R}/g)|E|^{2}$ and $(\Omega_{R}%
/g)|\Psi|^{2}$ are the photon and exciton numbers per unit area. Here,
$\Omega_{R}$ is the Rabi frequency and $g$ is the exciton-exciton interaction
constant. $\Delta=(\omega-\omega_{r})/\Omega_{R}$ describes detuning of the
pump frequency $\omega$ from the identical resonance frequencies of excitons
and cavity, $\omega_{r}$. The time $t$ is measured in units of $1/\Omega_{R}$.
$\gamma_{c}$ and $\gamma_{0}$ are the cavity and exciton damping constants
normalized to $\Omega_{R}$. The transverse coordinates $x$, $y$ are normalized
to the value $x_{0}=\sqrt{c/2k_{z}n\Omega_{R}}\sim 1\mu m$ where $c$ is the
vacuum light velocity, $n$ is the refractive index and $k_{z}=n\omega/c$ is
the wavenumber. The normalized amplitude of the external pump $E_0$ is
related to the physical incident intensity $I_{inc}$ as $|E_0|^{2}%
=g\gamma_{c}I_{inc}/\hbar\omega_{0}{\Omega_{R}}^{2}$
\cite{Wouters07}. $k_{0}$ is the transverse wavenumber of the pump,
with $k_{0}=0$ corresponding to normal incidence. As a guideline for
realistic estimates one can use parameters of a microcavity with a
single InGaAs/GaAs quantum well: $\hbar\Omega_{R}\simeq2.5$ meV,
$\hbar g\simeq10^{-4}$ eV$\mu$m$^{2}$, see
\cite{Baas2004,Carusotto2004}. Assuming the relaxation times of the
photonic and excitonic fields to be $2.5$ ps gives
$\gamma_{c,0}\simeq0.1$. In accordance with this set of parameters
the normalized driving amplitude $|E_0|^2=1$ corresponds to an
external pump intensity of about $10$ kW/cm$^{2}$. Optical
bistability appears for $|E_{p}|\sim0.1$ or about
$10^{2}-10^{3}$W/cm$^{2}$. $U(y)$ is the spatially varying shift of
the cavity resonance which represents a potential for photons. The
spatial inhomogeneities of the cavity and quantum wells are assumed
negligible, so that long range spatial effects can be observed. As,
for example, in the recent demonstration of the $~50\mu$m in
diameter anti-phase polariton $\pi$-state in a microcavity with
periodically patterned mirrors \cite{Lai2007}.

To understand the physics of CPS formation and to identify domains of
their existence we first  expand the linear polariton dispersion around the pump wavevector, and
arrive after an inverse Fourier transformation at a single  equation
with an effective nonlinearity and multiple dispersion terms. We
assume $U=0$ and look for the linear polariton eigenstates, i.e.,
the photons trapped in the pump free cavity ($E_0=0$, $\Delta
\rightarrow-\omega_{r}/\Omega_{R}$) coupled to noninteracting excitons.
Seeking solutions of Eqs. (1) in the form ${E,\Psi}={e_{k},\psi_{k}%
}e^{-i\Omega t+ikx}$ gives the eigenvalue problem:
\begin{equation}
i\Omega_{\pm}(k)\vec{p}_{k}=\left(
\begin{array}
[c]{cc}
ik^{2}+\gamma_{c}+i\omega_{r}/\Omega_{R} & -i\\
-i & \gamma_{0}+i\omega_{r}/\Omega_{R}
\end{array}
\right)  \vec{p}_{k}, \label{lin_waves}
\end{equation}
where $\vec{p}_{k}=\{e_{k},\psi_{k}\}$ is the polariton basis vector and
$\Omega_{\pm}(k)$ are the eigenfrequencies, corresponding to the upper- (+)
and lower- (-) branch of the polariton dispersion curve,
respectively, see Fig.~\ref{disp_ch}. The inflection point $\partial_{k}%
^{2}\Omega_{-}=0$ is located at $k=k_{d}$, where second order dispersion disappears.

The standard next step here would be to disregard the upper branch
and to consider the $k_{0}$ values close to zero, see, e.g.
\cite{Deveaud2007,Kavokin2008,Baas2004,Carusotto2004}. In order to
embrace the inflection point and to account for the varying
dispersion we modify this technique and include higher-order
dispersion terms. To achieve this we perform the Fourier transform
$\{{E,\Psi}\}\simeq\int{a(t,k)\vec{p}_{k}e^{ikx}}dk$, where
$a({t},k)$ is the Fourier amplitude of the $k$-th component and
$\vec{p}_{k}$ corresponds to the lower polariton branch. We assume
that the spectrum of the polariton wavepacket is centered around
$k_{0}$ and expand $\Omega_{-}(k)$ up to the fourth order in
$|k-k_{0}|$. The slowly varying amplitude of the polariton
wavepacket is then $A(t,x)=\int_{-\infty}^{\infty}a(t,k)e^{i(k-k_{0}%
)x}d(k-k_{0})$ and obeys:
\begin{align}
&  i\partial_{t}A+iD_{1}\partial_{x}A+D_{2}\partial_{x}^{2}A-iD_{3}%
\partial_{x}^{3}A-D_{4}\partial_{x}^{4}A-\nonumber\\
&  -\delta A-\xi\mid A\mid^{2}A=i\eta {E}_0, \label{LugLef_eq}%
\end{align}
where $\delta=\Omega_{-}(k_{0})-\omega/\Omega_{R}$, $\operatorname{Re}\delta$
is the polariton detuning from the pump frequency and $\operatorname{Im}%
\delta$ is the loss. $D_{1}=\partial_{k}\Omega_{-}|_{k_{0}}$ is the
transverse group velocity and
$D_{2}=(1/2)\partial_{k}^{2}\Omega_{-}|_{k_{0}}$,
$D_{3}=(1/6)\partial_{k}^{3}\Omega_{-}|_{k_{0}}$,
$D_{4}=(1/24)\partial _{k}^{4}\Omega_{-}|_{k_{0}}$ are the
dispersion coefficients [Fig.~\ref{disp_ch}(b)]. $\xi
=|\psi_{k_{0}}|^{4}/(|e_{k_{0}}|^{2}+|\psi_{k_{0}}|^{2})$ is the
effective nonlinearity and
$\eta=e_{k_{0}}/(|e_{k_{0}}|^{2}+|\psi_{k_{0})}|^{2})$.
For $k_{0}$ varying from zero
and towards, (but not too close to $k_{d}$)  $D_{2}>0$ dominates
over $D_{m>2}$ [Fig.~\ref{disp_ch}(b)]. In this case Eq. (\ref{LugLef_eq}) coincides with
the well-known weak-coupling Lugiato-Lefever (LL) model
\cite{LUGIATO1987}.

\begin{figure}[ptb]
\setlength{\epsfxsize}{2.5in}
\centerline{\epsfbox{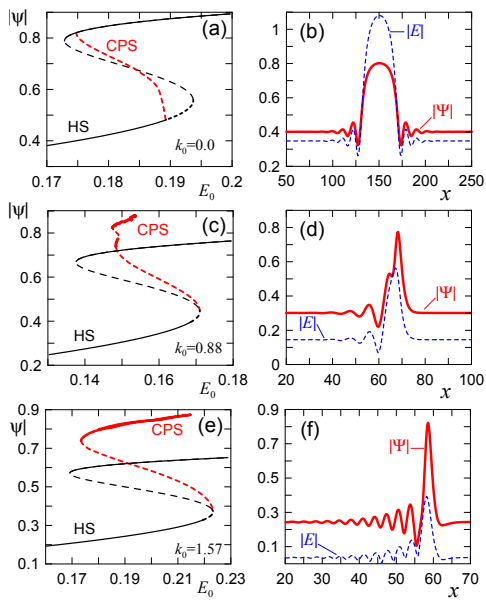}}\caption{(color online) Maxima of the
excitonic components of the solitons vs +
 (left column)
and the soliton profiles (right columns) for different $k_{0}$.
Black lines in the left column mark the peak amplitudes of the
spatially homogeneous solutions (HS) and the red ones of the
solitons. The full/dashed lines in the left column correspond to
stable/unstable solutions respectively. $E_0$ and $\Delta$ values
are: (a,b) $0.175$ and $-0.7$; (c,d) $0.149$ and $-0.39$; (e,f)
$0.19$ and $-0.05$
.}
\label{sol_branch}%
\end{figure}
In the LL models applied for the lower polaritons the dispersion is
always normal, $D_{2}>0$,  and the nonlinearity is repulsive, therefore
the only stable CPSs are the dark ones \cite{Yulin}, while bright solitons
exist, but are unstable [Fig. \ref{sol_branch}(a,b)]. However, $D_{2}(k)$ crosses zero at $k_{0}=k_{d}$ and
the $D_{3}$-term becomes the leading one around this point. Then the $D_{3}%
$-term drops and the polariton dispersion is determined by the
competing $D_{2}$ and $D_{4}$ terms. Both $D_{2}<0$ and $D_{4}>0$ favor the
existence of bright solitons for the repulsive nonlinearity and
therefore we expect to find them for $k_{0}>k_{d}$. Moreover, it is
evident that because of $D_{1}\neq0$ these solitons are forced to move.

While the above considerations are useful for  qualitative understanding, we
return to the original model (1) with $\partial_{y}=0$ in order to compute 1D
bright CPSs. In seeking for soliton solutions we have to take their motion into
account: $E(t,x)=\tilde{E}%
(\xi)e^{ik_{0}x}$, $\Psi(t,x)=\tilde{\Psi}(\xi)e^{ik_{0}x}$,
$\xi=x-vt$, where $v$ is the velocity (yet to be determined) close,
but not equal, to $D_{1}$, see Fig. 2(a). $\tilde{E}$ and
$\tilde{\Psi}$ obey
$(2k_{0}-v)\partial_{\xi}\tilde{E}-i\partial_{\xi}^{2}\tilde{E}+(\gamma
_{c}-i\Delta+ik_{0}^{2})\tilde{E}=i\tilde{\Psi}+E_0$ and
$-v\partial_{\xi}\tilde{\Psi}+(\gamma_{0}-i\Delta)\tilde{\Psi}+i|\tilde
{\Psi}|^{2}\tilde{\Psi}=i\tilde{E}$. Bright CPS solutions of the
above model are found by a modification of the Newton method
allowing to treat $v$ as an unknown variable. The maximum of $v$
occurs at $k=k_{d}$, then the soliton velocity decreases with
$k_{0}$ increasing and approaches zero, see Fig. 3(a). This is in
remarkable contrast to what happens if dispersion is parabolic,
where the soliton velocity continuously increases with $k_{0}$. For
our choice of parameters  $v$ must be multiplied by $4\times
10^6$m/s to give the physical velocity. It implies that the CPS with
$v=0.25$ will traverse across the typical distance of $100\mu$m in
$100$ps.  This is 40 times larger than the polariton life time.
Hence, once excited, the solitons  have enough time to shape and be
observed during their motion.

Plots illustrating the
dependence of the maximal soliton amplitude from the pump field and the
associated bistability curves for the homogeneous ($\partial_{x,y}=0$) solutions are shown in the
left column of Fig. 2. The right column shows the corresponding transverse
soliton profiles. For $k_{0}=0$ we are reproducing the results of \cite{Yulin}
with  bright CPSs being unstable. For $k_{0}$ just below and above $k_{d}$ the
branch of bright CPSs starts bending and islands of their stability start to
emerge and expand with $k_{0}$ increasing.
\begin{figure}[ptb]
\setlength{\epsfxsize}{2.5in} \centerline{\epsfbox{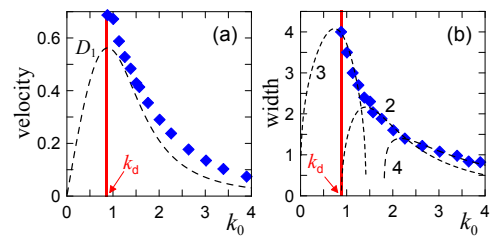}}\caption{ (color
online) (a) Diamonds show numerically computed soliton velocity vs. $k_{0}$.
Dashed line shows $D_{1}(k_{0})$. (b) Diamonds show the soliton width vs
$k_{0}$. Dashed lines show the appropriately scaled
$\sqrt[m]{D_{m}}$, where $m=2,3,4$. }%
\label{sol_width}%
\end{figure}

To reveal the influence of different dispersion orders on the CPSs we plotted
the dependence of the CPS width on the inclination parameter $k_{0}$ in
Fig.~\ref{sol_width}(b). Since not only the dispersion coefficients vary with
$k_{0}$, but also $\operatorname{Re}\delta$ (which  strongly
influences the CPS width), we have kept the latter fixed at $\operatorname{Re}%
\delta=-0.3$, by adjusting the detuning $\Delta$ respectively. For large
$k_{0}$ the excitonic part of CPSs dominates over the photonic component, see
Fig. 2. That is why the CPSs can be much narrower than the pure photonic
cavity solitons in the weak coupling regime and may attain widths well below
the ones allowed by the photonic dispersion, although this happens at the
price of oscillatory tails. The CPS width is expected to scale with the
dominant dispersion coefficients as $\sim\sqrt[m]{D_{m}}$ ($m=2,3,4$).
Fig.~\ref{sol_width}(b) compares the numerically found soliton width
(diamonds) with the scaling given by the different dispersion orders (dashed
lines). Third-order dispersion describes well the soliton width for
relatively small $k_{0}$ where $D_{2}\simeq0$. Further inclination brings
$D_{2}$ on the top, while the third-order dispersion vanishes. A further
increase of $k_{0}$ brings fourth-order dispersion into  play, which starts
to compete with $D_{2}$.
\begin{figure}[ptb]
\setlength{\epsfxsize}{2.5 in}
\centerline{\epsfbox{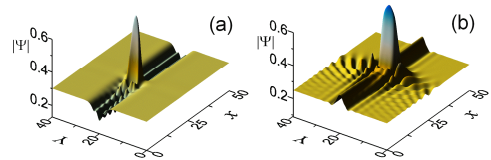}}\caption{ (color online)
 (a) Soliton moving along the potential
attractive for linear polaritons: $U=b\exp[-(y/2)^{8}]$, $b=2$,
$E_{0}=0.18$. Other parameters as in Fig. 2(e,f).
(b) Soliton moving along the repulsive potential $b=-1$, $E_{0}=0.213$.}%
\label{2d_prof}%
\end{figure}

In order to use CPSs as information bits, it is desirable to have them
robustly localized in both transverse dimensions. However in the homogeneous cavity
this is impossible to achieve, since the signs of
polariton dispersion  are opposite in orthogonal directions.
To solve this problem  we propose to shift the cavity
detuning in such a way that outside the waveguide area the soliton
existence conditions are not satisfied. This is particularly easy to
achieve if the operating frequency is not far from the boundary of
the existence domain. With this approach, it does not
matter if one uses $U(y)$, which traps  linear waves, or
$U(y)$, which repels them. Thus using of strongly nonlinear polariton waves
extends the list of the possible trapping techniques already explored for linear cavity
polariton \cite{Balili2007,Liew2008,Lai2007}. Fig. \ref{2d_prof}(a) demonstrates
soliton guiding in the trapping potential and Fig. \ref{2d_prof}(b)
in the repelling potential. The trapping potential looks
preferential for the soliton guidance, because it reduces soliton
interaction with the quasi-linear waves. The latter are far more
pronounced in the case of the repelling potential, see Fig.
\ref{2d_prof}(b). Practical excitation of CPSs is no different from their
weakly coupled counterparts and requires short term
application of the addressing beam in addition to the pump $E_0$
\cite{Barland2002,Pedaci2008}.
\begin{figure}[ptb]
\setlength{\epsfxsize}{2.0in}
\centerline{\epsfbox{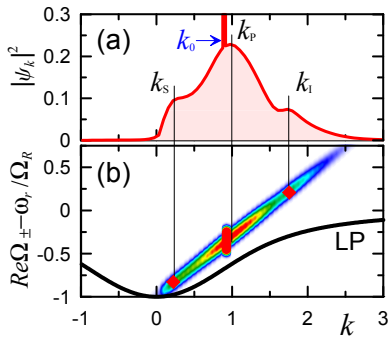}}\caption{(color online) (a) Excitonic
part of the soliton spectrum ($|\psi_k|$) of the the soliton shown
in Fig. 2(d).
 (b) The density plot is the 2D generalization of $|\psi_k|$ from (a).
 The slope of this plot shows
 dependence of the soliton frequency on $k$. The full line is LP dispersion.}
\label{disp_sol}
\end{figure}

In conclusion, we have shown that stable moving 1D cavity polariton
solitons exist providing the lower polaritons are
used near or beyond the inflection point of their dispersion. We have also demonstrated the
continuous transformation between cavity soliton polaritons shaped
by the different dispersion orders, including the regimes where the
influence of the usual second order dispersion is negligible. We
have proposed a mechanism for their guiding and localization in the
second transverse dimension. Our results
form the basis for future studies of
soliton-polariton logic and processing schemes.

{\em Note.} During the revision of this manuscript the related
experimental observations of moving quasi-localised polaritons have
been published \cite{amo}. Brief preliminary comparison of
dispersion plots from \cite{amo}  with our results is very
encouraging. To reveal the soliton frequency content we take the
Fourier transform of its excitonic component:
$\psi_k=\int_{-\infty}^{\infty} \tilde\Psi(x-vt)e^{ik_0x+ikx}dx=
f(k-k_0)e^{ivt(k_0-k)}$, where $f(k)=\int_{-\infty}^{\infty}
\tilde\Psi(\xi)e^{ik\xi}d\xi$. $|\psi_k|$ is plotted in Fig. 5(a).
Apart from the infinite peak at the at $k_0$ corresponding to the
background, the soliton spectrum proper shows 3-peak structure, such
that $2k_p=k_i+k_s$, where $k_p\approx k_0\approx k_d$. The  idler
field $k_i$ was externally seeded and the signal $k_s$ was generated
in  Ref. \cite{amo}, while  both of them are the integral parts of
our soliton solution. Soliton frequency calculated from $\psi_k$ is
a straight line shifted upwards from the dispersion of linear
polaritons (Fig. 5(b)). Figs. 5(a,b) essentially match the
corresponding data from \cite{amo}.


\begin{thebibliography}{99}                                                                                               %


\bibitem {Deveaud2007}B.~Deveaud, editor.
\newblock {\em The Physics of Semiconductor Microcavities}
\newblock (Willey-VCH, 2007).

\bibitem {Kavokin2008}A.~Kavokin, J.J. Baumberg, G.~Malpuech, and F.P. Laussy.
\newblock {\em Microcavities} \newblock (Oxford University Press, 2007).

\bibitem {Peschel2003}U.~Peschel \emph{et al.}, IEEE J. Quant. Electr.
\textbf{39}, 51 (2003).

\bibitem {Tredicucci1996}A.~Tredicucci \emph{et al.}, Phys. Rev. A
\textbf{54}, 3493 (1996).

\bibitem {Baas2004}A.~Baas \emph{et al.}, Phys. Rev. A \textbf{69}, 023809 (2004).

\bibitem {Carusotto2004}I.~Carusotto and C.~Ciuti,
Phys. Rev. Lett. \textbf{93}, 166401 (2004).

\bibitem {cnrs}D. Bajoni \emph{et al.}, Phys. Rev. Lett. \textbf{101}, 266402 (2008).

\bibitem {Gippius2004}N.~A. Gippius \emph{et al.}, Europhys. Lett.
\textbf{67}, 997 (2004).

\bibitem {Krizhanovskii2008}D.~N. Krizhanovskii \emph{et al.}, Phys. Rev. B
\textbf{77}, 115336 (2008).

\bibitem {Houdre2000}R.~Houdre \emph{et al.}, Phys. Rev. Lett. \textbf{85},
2793 (2000).

\bibitem {Savvidis2000}P.~G. Savvidis \emph{et al.}, Phys. Rev. Lett.
\textbf{84}, 1547 (2000).

\bibitem {Diederichs2006}C.~Diederichs \emph{et al.}, Nature \textbf{440}, 904 (2006).

\bibitem {Wouters07}M. Wouters and I. Carusotto, Phys. Rev. B \textbf{75},
075332 (2007).

\bibitem {Stevenson2000}R.~M. Stevenson \emph{et al.}, Phys. Rev. Lett.
\textbf{85}, 3680 (2000).

\bibitem {Deng2002}H.~Deng \emph{et al.}, Science \textbf{298}, 199 (2002).

\bibitem {Kasprzak2006}J.~Kasprzak \emph{et al.}, Nature \textbf{443}, 409 (2006).

\bibitem {Balili2007}R.~Balili \emph{et al.}, Science \textbf{316}, 1007 (2007).

\bibitem {Love2008}A.~P.~D. Love \emph{et al.}, Phys. Rev. Lett. \textbf{101},
067404 (2008).

\bibitem {Baas2008}A.~Baas \emph{et al.},
Phys. Rev. Lett. \textbf{100}, 170401 (2008).



\bibitem {Eisenberg2000}H.~S. Eisenberg \emph{et al.},
Phys. Rev. Lett. \textbf{85}, 1863 (2000).

\bibitem {Pertsch2002}T.~Pertsch \emph{et al.},
Phys. Rev. Lett. \textbf{88}, 093901 (2002).



\bibitem {Greiner2002}
B. Eiermann \emph{et al.}, Phys. Rev. Lett. \textbf{91}, 060402 (2003); T.
Anker \emph{et al.}, Phys. Rev. Lett. \textbf{94}, 020403 (2005).

\bibitem {Ebbesen1998}T.~W. Ebbesen \emph{et al.},
Nature \textbf{415}, 667 (1998).

\bibitem {Lai2007}C.~W. Lai \emph{et al.},
Nature \textbf{450}, 529 (2007).

\bibitem {Cho2005}K.~Cho \emph{et al.},
Phys. Rev. Lett. \textbf{94}, 226406 (2005).

\bibitem {Fleischer2003}J.~W. Fleischer \emph{et al.},
Nature \textbf{422}, 147 (2003).

\bibitem {Kivshar2001}Y.S. Kivshar and G.~Agrawal.
\newblock {\em Optical Solitons: From Fibers to Photonic Crystals}
\newblock (Academic Press, 2001).

\bibitem {Lederer2008}F. Lederer \emph{et al.}, Phys. Rep. {\bf 463}, 1  (2008).



\bibitem {Staliunas2003}
K.~Staliunas, Phys. Rev. Lett. \textbf{91}, 053901 (2003);
D.~Gomila, R.~Zambrini, and G.~L. Oppo, Phys. Rev. Lett.
\textbf{92}, 253904 (2004); E.A. Ultanir {\em et al.}, Opt. Lett.
{\bf 29},  845 (2004); A.~V. Yulin, D.~V. Skryabin, and P.~S.~J.
Russell, Opt. Exp. \textbf{13}, 3529 (2005); O.~Egorov, F.~Lederer,
and K.~Staliunas, Opt. Lett. \textbf{32}, 2106 (2007); O.~Egorov and
F.~Lederer, Opt. Exp. \textbf{16}, 6050 (2008); K.~Staliunas
\emph{et al.}, Phys. Rev. Lett. \textbf{101}, 153903 (2008).



\bibitem {Brambilla1997}M.~Brambilla \emph{et al.}, Phys. Rev. Lett.
\textbf{79}, 2042 (1997).

\bibitem {Barland2002}S.~Barland \emph{et al.}, Nature \textbf{419}, 699 (2002).

\bibitem {Pedaci2008}F.~Pedaci \emph{et al.}, Appl. Phys. Lett. \textbf{92},
011101 (2008).





\bibitem {Yulin}A.~Yulin \emph{et al.}, Phys. Rev. A \textbf{78}, 061801(R) (2008).

\bibitem {Larionova2008}Y.~Larionova \emph{et al.},  Opt. Lett.
\textbf{33}, 321 (2008).

\bibitem {LUGIATO1987}L.A.Lugiato and R.Lefever, Phys. Rev. Lett. \textbf{58},
2209 (1987).

\bibitem {Liew2008}T.~C.~H. Liew \emph{et al.}, Phys. Rev. Lett. \textbf{101},
016402 (2008).

\bibitem{amo}
A. Amo {\em et al.}, Nature \textbf{457}, 291 (2009).

\end{thebibliography}
\end{document}